 \documentclass[final,5p,times,twocolumn]{elsarticle}

\usepackage{graphicx}
\usepackage{amssymb}
\usepackage{multirow}

\journal{Elsevier}

\begin{document}

\begin{frontmatter}



\title{How people make friends in social networking sites--A microscopic perspective}
\author[focal]{Haibo Hu}
\author[rvt]{Xiaofan Wang}

%\cortext[cor1]{Corresponding author. Tel.: +86 21 64252567}

\address[focal]{School of Business, East China University of Science and
Technology, Shanghai 200237, China}
\address[rvt]{Complex Networks and Control Lab, Shanghai Jiao Tong University, Shanghai 200240,
China}

\begin{abstract}
We study the detailed growth of a social networking site with full
temporal information by examining the creation process of each
friendship relation that can collectively lead to the macroscopic
properties of the network. We first study the reciprocal behavior of
users, and find that link requests are quickly responded to and that
the distribution of reciprocation intervals decays in an exponential
form. The degrees of inviters/accepters are slightly negatively
correlative with reciprocation time. In addition, the temporal
feature of the online community shows that the distributions of
intervals of user behaviors, such as sending or accepting link
requests, follow a power law with a universal exponent, and peaks
emerge for intervals of an integral day. We finally study the
preferential selection and linking phenomena of the social
networking site and find that, for the former, a linear preference
holds for preferential sending and reception, and for the latter, a
linear preference also holds for preferential acceptance, creation,
and attachment. Based on the linearly preferential linking, we put
forward an analyzable network model which can reproduce the degree
distribution of the network. The research framework presented in the
paper could provide a potential insight into how the micro-motives
of users lead to the global structure of online social networks.
\end{abstract}


\end{frontmatter}


%\newpage
\section{Introduction}

At present the World Wide Web (WWW) is undergoing a landmark
revolution from the traditional Web 1.0 to Web 2.0 characterized by
social collaborative technologies, such as social networking sites
(SNSs), blogs, Wiki, and folksonomy [1]. As a fast growing business,
many SNSs of different scopes and purposes have emerged in the
Internet, many of which, such as \emph{MySpace} [2, 3],
\emph{Facebook} [4-7], and \emph{Orkut} [2, 8], are among the most
popular sites on the Web according to Alexa.com. Users of these
sites, by establishing friendship relations with other users, can
form online social networks (OSNs), which provide an online private
space for individuals and tools for interacting with other people
over the Internet. Both the popularity of these sites and
availability of network data sets offer a unique opportunity to
study the dynamics of OSNs at scale. It is believed that having a
proper understanding of how OSNs evolve can provide insights into
the network structure, allow predictions of future growth, and
enable exploration of human behaviors on networks [9-13].

Recently, the structure and evolution of OSNs have been extensively
investigated by scholars of diverse disciplines. Golder \emph{et
al.} studied the structural properties of \emph{Facebook} and found
that the tail of its degree distribution is a power law which is
different from the traditional exponential distribution of real-life
social networks [4]. However, a mean of 179.53 friends per user for
\emph{Facebook} [4] or a mean of 137.1 friends per user for
\emph{MySpace} [2] is close to Dunbar's number of 150, which is a
limit on the number of manageable relations by human based on their
neocortex size [14]. Holme \emph{et al.} studied the structural
evolution of \emph{Pussokram} and found that its degree correlation
coefficient is always negative over time, i.e. disassortative mixing
[15], which is in stark contrast to the significant assortative
mixing for real-world social networks [16]. Viswanath \emph{et al.}
studied the structural evolution of the activity network of
\emph{Facebook} and found that the average degree, clustering
coefficient, and average path length are all relatively stable over
time [6]. Hu \& Wang studied the evolution of \emph{Wealink} [17,
18] and found that many network properties show obvious non-monotone
feature, including a sigmoid growth of network scale which was also
observed by Chun \emph{et al.} in \emph{Cyworld} [19], and a
transition from degree assortativity characteristic of real social
networks to degree disassortativity characteristic of many OSNs
which was also observed by Szell \& Thurner in \emph{Pardus} [20].

Despite the advancement, we find that to date most research on OSNs
has focused on either the structural properties of a certain
snapshot of networks or the multi-snapshots of evolving networks
rather than detailed microscopic growth dynamics. For the research
framework of network evolution from a macroscopic viewpoint it is
usually hard to reveal underlying mechanisms and growth processes
governing the large-scale features of the observed network
structure. In this paper, to gain better insight into the growth of
networks, based on empirical data, we study the detailed process of
people making friends in an OSN from a microscopic point of view.
Instead of investigating the global network structure or structural
metric evolution, we focus directly on the microscopic user
behaviors per se, i.e., we study the properties of a sequence of the
arrivals of each edge or the formations of each friend relation.

\section{Data set}

In this paper, we will focus on \emph{Wealink}, a large SNS in China
whose users are mostly professionals, typically businessmen and
office clerks [17, 18]. Each registered user has a profile,
including his/her list of friends. If we view the users as nodes $V$
and friend relations as edges $E$, an undirected friendship network
$G$($V$, $E$) can be constructed from \emph{Wealink}. For privacy
reasons, the data, logged from 0:00:00 h on 11 May 2005 (the
inception day for the Internet community) to 15:23:42 h on 22 August
2007, include only each user's ID and list of friends, and the time
of sending link invitations and accepting requests for each friend
relation.

The finial data format, as shown in Fig. 1, is a time-ordered list
of triples $<$$From$, $To$, $When$$>$. For instance, $<$$U_1$,
$U_2$, $T_1$$>$ indicates that, at time $T_1$, user $U_1$ sends a
link request to user $U_2$, i.e., sends a friendship invitation to
$U_2$, while $<$$U_2$, $U_1$, $T_6$$>$ indicates that, at $T_6$,
$U_2$ accepts $U_1$'s request and they become friends, i.e., a new
edge connecting $U_2$ and $U_1$ appears in the OSN. Thus only when
the sent invitations are accepted will the friend relations or
network links be established. The online community is a dynamically
evolving one with new users joining the community and new
connections established between users.

\begin{figure*}
  \centerline{\includegraphics[height=1.8in]{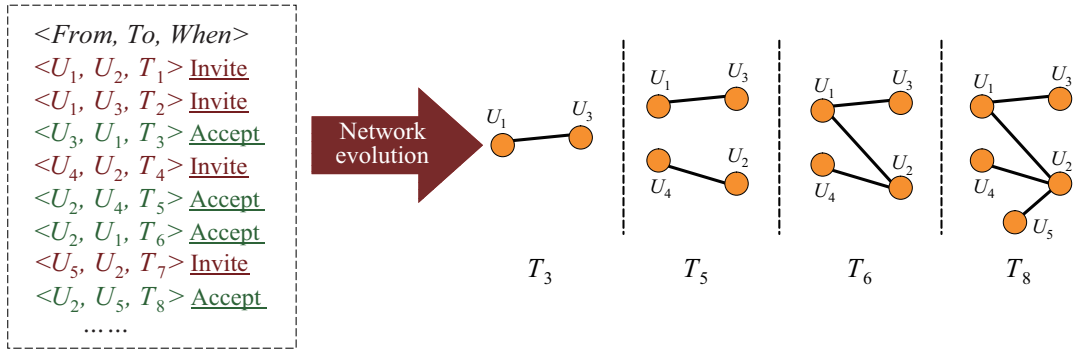}}
  \caption{Data format and evolution of OSN \emph{Wealink}.}
\end{figure*}

\section{Reciprocal behavior}

Like most OSNs, in \emph{Wealink}, a user invites another user to be
his/her friend; if the invited user accepts the invitation, a friend
relation is established between them and a new edge connecting them
appears (see Fig. 1). Thus the friendship is constructed by
bilateral agreement. The degree of a user, i.e., the number of
friends, will appear on his/her profile, which can be browsed by all
the other users. During our data collection period, 273 209 sent
link requests have been accepted and only 186 ones have not yet been
accepted. Thus, in the following analysis, we will focus on the 273
209 sent link requests and their corresponding accepted ones with
full temporal information.

We first scrutinize the reciprocation of users, i.e., the sending of
a link request from one user to another (as happens at $T_1$ in Fig.
1) causes following acceptance of the request ($T_6$). Fig. 2(a)
shows the complementary cumulative distribution of the intervals
between sending and accepting link requests in \emph{Wealink}. It is
clear that users often quickly responded to link requests and
reciprocated them. The interval distribution decays approximately
exponentially. The least squares fitting gives ${P_{\rm{c}}}(t) \sim
{{\rm{e}}^{ - 0.011t}}$ with ${{\rm{R}}^2} = {\rm{0}}{\rm{.958}}$.
In fact, as shown in Fig. 2(b), 67.04\% of all reciprocal behavior
occurred within one day (24 hours) after the initial link requests
and 84.25\% of sent link requests were accepted within one month (30
days). \emph{Wealink} informs users by email of new incoming link
requests. It is quite possible that many users reciprocated requests
as a matter of courtesy and respect.

\begin{figure*}
  \centerline{\includegraphics[height=2.5in]{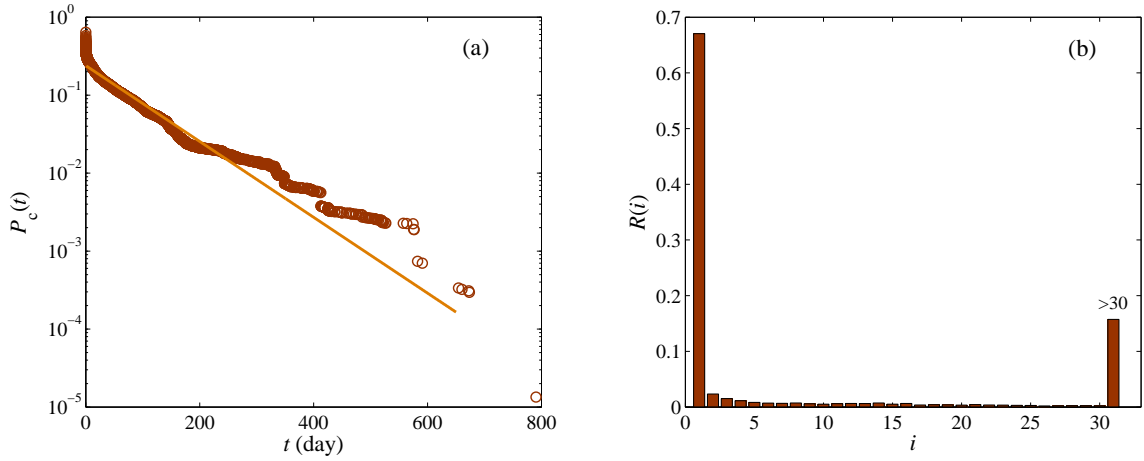}}
  \caption{(a) The complementary cumulative distribution of time
intervals between sending and
  accepting invitations. The solid line represents an exponential distribution fit.
(b) The ratio of sent link requests which were accepted on the $i$th
day after the initial invitations.}
\end{figure*}

Recently inspired by the pioneering works of Barab\'{a}si \emph{et
al}., there has been increasing interest for physicists and computer
scientists in the research of human dynamics [21, 22], which focuses
on the time interval distribution between two consecutive actions
performed by individuals. The examples of such temporal statistics
include the inter-event time distribution between two consecutive
emails sent out by users, two consecutive visits to a web portal by
users, and two consecutive library loans made by individuals.
Empirical studies have shown that many distributions of inter-event
time follow a power law. However, the exponential reciprocation
interval distribution is in distinct contrast to the power law
distribution of waiting time in emails (i.e., the time taken by
users to reply to received emails). The importance of different
emails is different. A reasonable hypothesis is that there can be
correlation between the importance of emails and reputation/status
of senders or ``social closeness" to senders. Thus users can reply
to received emails based on some perceived priority, and the timing
of the replies will be heavy tailed. In contrast, there is no
obvious priority for the reciprocal behavior of users in OSNs; thus
an exponential distribution will well characterize the reciprocation
interval distribution.

An interesting question is whether the users tended to reciprocate
incoming link requests quickly regardless of how many friends the
inviters or accepters had. To answer the question, we study the
correlation between reciprocation time and the degrees of
inviters/accepters at the time of sending link requests. Fig. 3(a)
shows the density plot based on hexagonal binning for the relation
between degrees of inviters $k$ and reciprocation time $t$, where
the cases with small $k$ and $t$ dominate. The Pearson correlation
coefficient between $k$ and $t$ is -0.02, indicating slightly
negative correlation. Fig. 3(b) shows the relation between $k$ and
mean reciprocation time $\left\langle t \right\rangle$ with
logarithmic binning and error bars. $\left\langle t \right\rangle$
exhibits mild descending trend as $k$ increases. Fig. 4 shows the
relation between degrees of accepters $k$ and reciprocation time
$t$, which is similar to that shown in Fig. 3. The Pearson
correlation coefficient between $k$ and $t$ is -0.05, also
indicating mildly negative correlation.

\begin{figure*}
  \centerline{\includegraphics[height=2.5in]{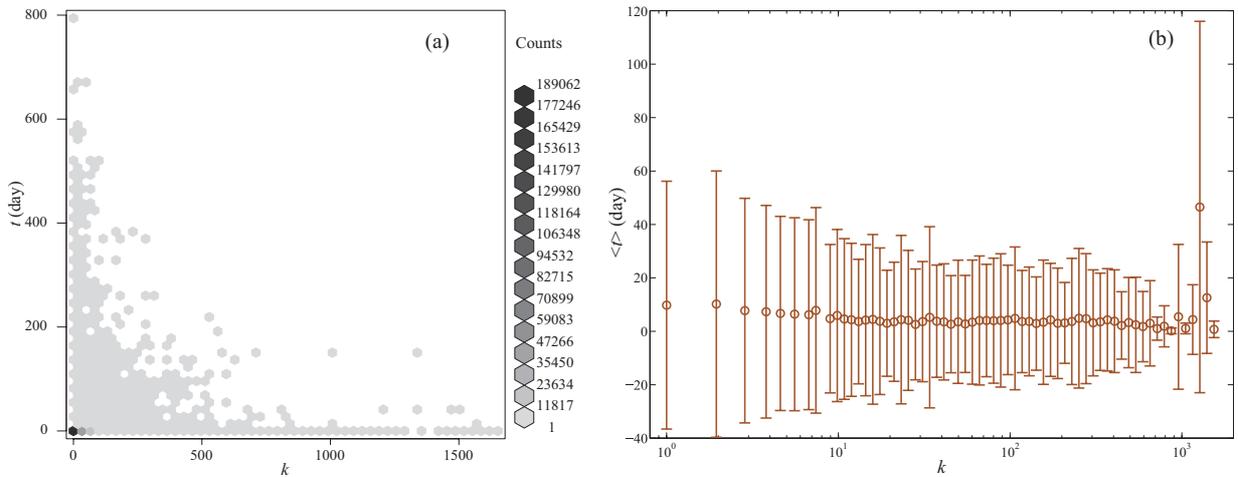}}
  \caption{Relation between degrees of inviters $k$ and reciprocation
time $t$. (a) Density plot based on hexagonal binning. (b) Relation
between $k$ and mean reciprocation time $\left\langle  t
\right\rangle$ with logarithmic binning. Error bars with $ \pm
{\rm{1}}$ standard deviation are also shown.}
\end{figure*}

\begin{figure*}
  \centerline{\includegraphics[height=2.5in]{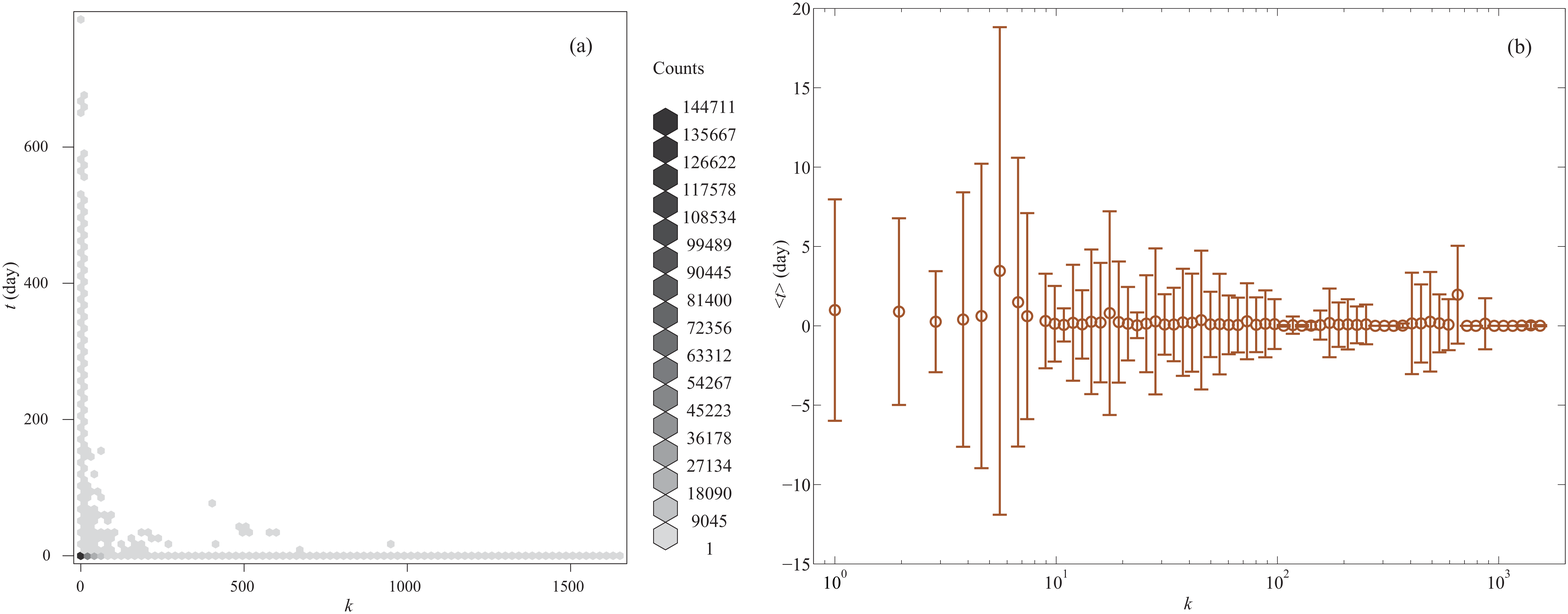}}
  \caption{Relation between degrees of accepters $k$ and reciprocation
time $t$. (a) Density plot based on hexagonal binning. (b) Relation
between $k$ and mean reciprocation time $\left\langle  t
\right\rangle$ with logarithmic binning. Error bars with $ \pm
{\rm{1}}$ standard deviation are also shown.}
\end{figure*}

\section{Types of users and edges}

The total number of users during the data collection period is
$N$=223 482. Obviously the users can be divided into three classes:
active users who sent link requests but have never received ones,
passive users who received requests but have never sent ones, and
mixed users who both sent and received requests. As shown in Tab. 1,
we find that most users belong to the former two classes. For the
very popular SNSs, such as \emph{Facebook} and \emph{MySpace}, due
to the high activity and degree values of users, most users could be
mixed ones. However, \emph{Wealink} is a very professional OSN with
a mean degree of only 2.53. The activity of most users is low; after
joining in the OSN they either send link requests to a few old users
(acquaintance in real life very likely) or receive link invitations
from several old users. Among the mixed users, there exists obvious
positive correlation between the numbers of times of sending
invitations and accepting invitations, and the Pearson correlation
coefficient is 0.48. As shown in Fig. 5, we find that the more link
requests a user sends, the more requests he/she will receive, and
vice versa.

\begin{table}
\caption{The numbers of users of different types.} \center
\begin{tabular}{llll}
\hline Type  & Active & Mixed   &   Passive\\  \hline
Number & 128 589 & 16 060 & 78 833\\
Percentage & 57.54\% & 7.19\% & 35.27\% \\
\hline
\end{tabular}
\center
\end{table}

\begin{figure*}
  \centerline{\includegraphics[height=4.5in]{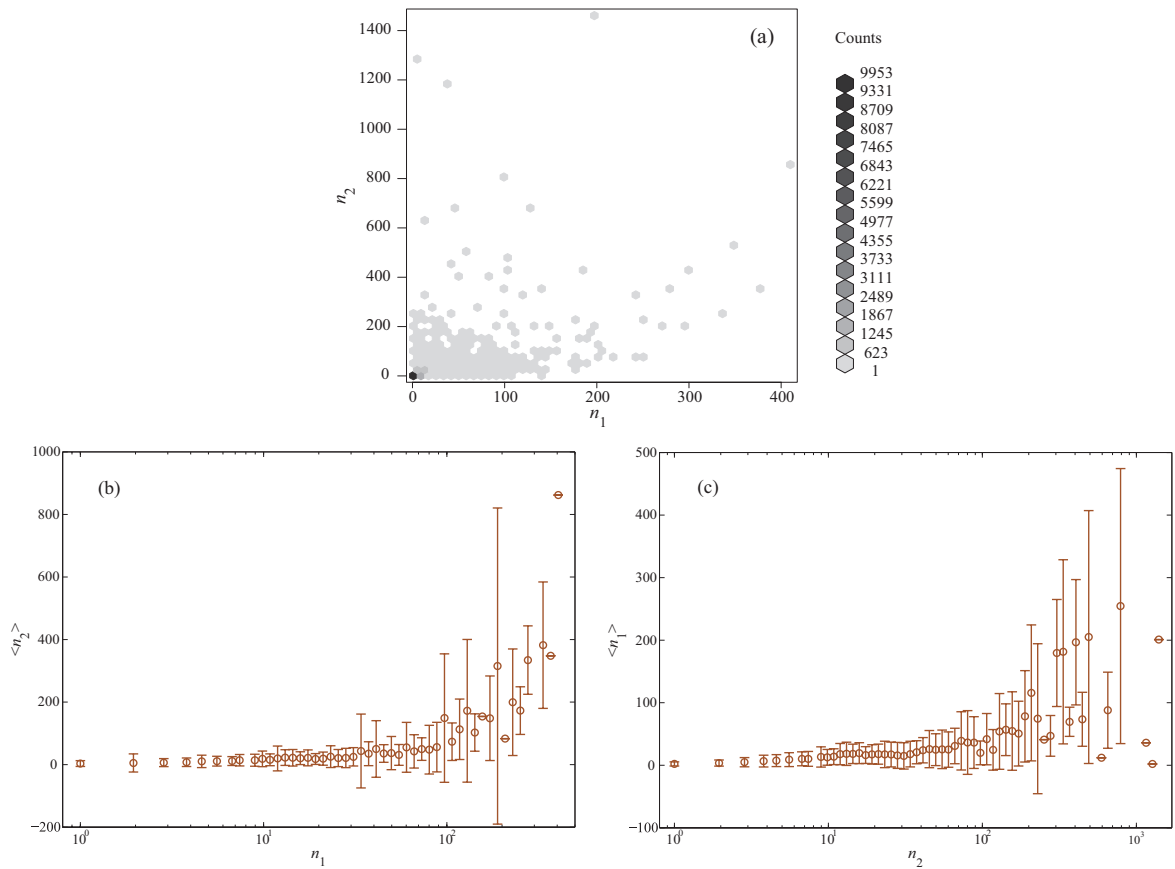}}
  \caption{Correlation between the numbers of times of sending
invitations $n_1$ and accepting invitations $n_2$. (a) Density plot
based on hexagonal binning. (b) Relation between $n_1$ and
$\left\langle  n_2 \right\rangle$ with logarithmic binning. (c)
Relation between $n_2$ and $\left\langle  n_1 \right\rangle$ with
logarithmic binning. Error bars with $ \pm {\rm{1}}$ standard
deviation are also shown.}
\end{figure*}

The finial network density is only ${\rm{1}}{\rm{.09}} \times
10^{{\rm{ - 5}}} $, and what results in the sparseness? As shown in
Tab. 2, the $E$=273 209 edges can be divided into four classes, and
$A$$-$$B$ type expresses that initially $A$ users sent link requests
to $B$ users. ``Old" means that the users have been in the network;
they joined in the network some time ago and they either have sent
at least one link request to other users or have received at least
one link request from other users. ``New" means that the users have
joined in the network; however, they neither have sent link requests
to other users nor have they received link requests from other
users. It is shown that in \emph{Wealink} most links are established
by old users sending requests to new users (more than 30\%) and new
users sending requests to old users (approximately 50\%). The number
of edges of Old-Old type is relatively small, leading to the
sparseness of the network.

\begin{table}
\caption{The numbers of edges of different types.} \center
\begin{tabular}{lllll}
\hline Type  & Old-Old & Old-New & New-Old & New-New\\  \hline
Number & 52 980 & 82 740 & 134 236 & 3 253 \\
Percentage & 19.39\% & 30.28\% & 49.13\% & 1.19\% \\
\hline
\end{tabular}
\center
\end{table}

\section{Temporal characteristics of linking}

We study the time interval distribution between two link events. As
shown in Fig. 6, the distributions of intervals between consecutive
sending link requests (i.e., between $T_1$ and $T_2$, $T_2$ and
$T_4$, and so on in Fig. 1), accepting requests (i.e., between $T_3$
and $T_5$, $T_5$ and $T_6$, and so on in Fig. 1) and any two events
(i.e. between $T_i$ and $T_{i+1}$ ($i \ge 1$) in Fig. 1) all follow
a power law with a universal exponent 1.89, which diverges from the
exponential distribution predicted by a traditional Poisson process
and indicates bursts of rapidly occurring events separated by long
periods of inactivity. Several peaks appear for intervals of an
integral day in the tails of the distributions, indicating the daily
periodicity corresponding to human life habits.

\begin{figure*}
  \centerline{\includegraphics[height=2.5in]{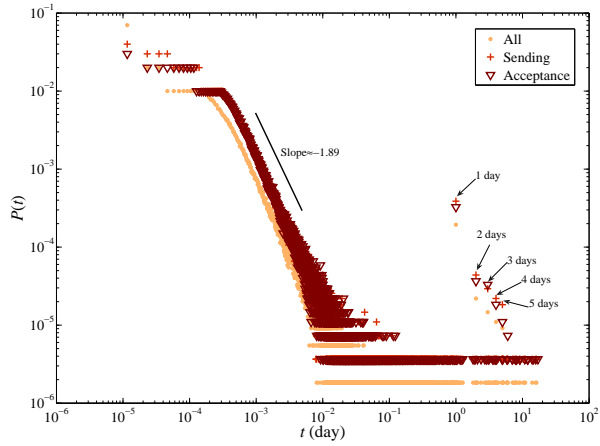}}
  \caption{Temporal characteristics of link request sending and
acceptance.}
\end{figure*}

\section{Preferential selection}

Preferential selection means that, for a time-ordered list of
individual appearance, the more likely an individual appeared
before, the more possibly the individual will occur once again. We
separate the preferential selection into two aspects: preferential
sending and preferential reception. Preferential sending describes
the mechanism by which users send new link requests with probability
proportional to some power of the numbers of their sent link
invitations before, and preferential reception describes the
mechanism by which users receive new link requests with probability
proportional to some power of the numbers of their received link
invitations before.

Fig. 7 presents the schematic illustration of sending and reception
sequences of OSNs. The former is a time-ordered list of users
sending link invitations, and the latter is a time-ordered list of
users receiving link invitations. In both sequences, the more
frequently a user appeared before, the more likely the user will
occur once again.

\begin{figure*}
  \centerline{\includegraphics[height=1.8in]{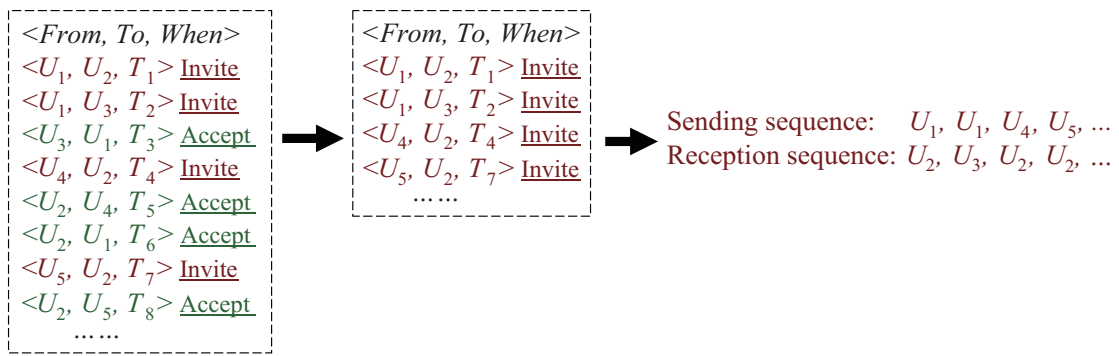}}
  \caption{A schematic illustration of sending and reception sequences
for \emph{Wealink}.}
\end{figure*}

Let $k_i$ be the number of sent or received link invitations for
user $i$. The probability that user $i$ with frequency $k_i$ is
chosen to send or receive a link request once again can be expressed
as
\begin{equation}
  \prod (k_i ) = \frac{{k_i^\beta  }}{{\sum\nolimits_j {k_j^\beta  }
  }}.
  \label{1}
\end{equation}
Thus we can compute the probability $\prod (k)$ that an old user of
frequency $k$ is chosen, and it is normalized by the number of users
of frequency $k$ that exist just before this step [23, 24]:
\begin{equation}
  \prod (k) = \frac{{\sum\nolimits_t {\left[ {e_t  = v \wedge k_v (t - 1) = k} \right]} }}{{\sum\nolimits_t {\left| {\left\{ {u:k_u (t - 1) = k} \right\}} \right|} }} \sim
  k^\beta,
  \label{2}
\end{equation}
where $e_t  = v \wedge k_v (t - 1) = k$ represents that at time $t$
the old user whose frequency is $k$ at time $t-1$ is chosen. We use
$[ \cdot ]$ to denote a predicate (which takes a value of 1 if the
expression is true, else 0). Generally, $\prod (k)$ has significant
fluctuations, particularly for large $k$. To reduce the noise level,
instead of $\prod (k)$, we study the cumulative function:
\begin{equation}
  \kappa (k) = \int_0^k {\prod (k){\rm{d}}k}  \sim k^{\beta  + 1}.
  \label{3}
\end{equation}

Fig. 8 shows how the frequency $k$ of users is related to the
preference metric $\kappa$. $\beta  \approx 1$ for both preferential
sending and preferential reception, indicating linear preference.

\begin{figure*}
  \centerline{\includegraphics[height=2.5in]{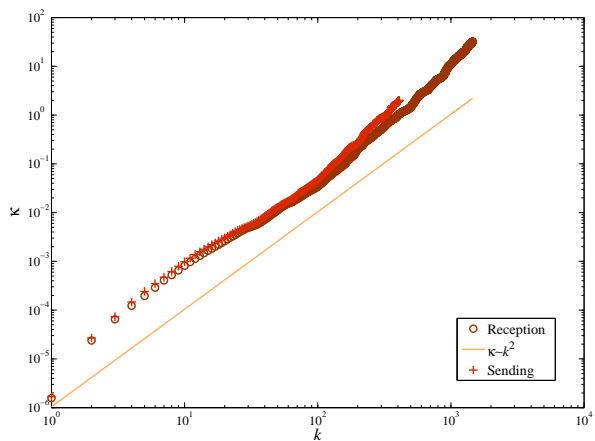}}
  \caption{Testing preferential selection for users of sending and
receiving invitations in \emph{Wealink}.}
\end{figure*}

It is natural that, in the sending or reception sequence, the number
of distinct users $N$ increases with sequence length $T$. Fig. 9
shows the growth pattern of $N$ with $T$ for \emph{Wealink}. $N
\propto T$, indicating that the appearance probability of new users
is a constant, $\alpha=N/T$. According to the Simon model [25],
based on linear preferential selection and constant appearance
probability of new users, the complementary cumulative distributions
of the numbers of sent invitations and received invitations for the
users of \emph{Wealink} follow a power law $P_{\rm{c}} (n) \sim n^{
- \left( {\frac{{\rm{1}}}{{{\rm{1 - }}\alpha }}} \right)} $. Based
on empirical data, for the inviters, we obtain $\alpha  =
{\rm{0}}{\rm{.53}}$ and $P_{\rm{c}} (n) \sim n^{{\rm{ -
2}}{\rm{.13}}} $, and for the receivers, $\alpha  =
{\rm{0}}{\rm{.35}}$ and $P_{\rm{c}} (n) \sim n^{{\rm{ -
1}}{\rm{.54}}} $. Fig. 10 shows the distribution functions of the
frequencies of inviters and receivers, and the tails of both
distributions show power law behavior. The power law exponents
achieve proper agreement with the predicted values of the Simon
model, $1/(1-\alpha)$.

\begin{figure*}
  \centerline{\includegraphics[height=2.5in]{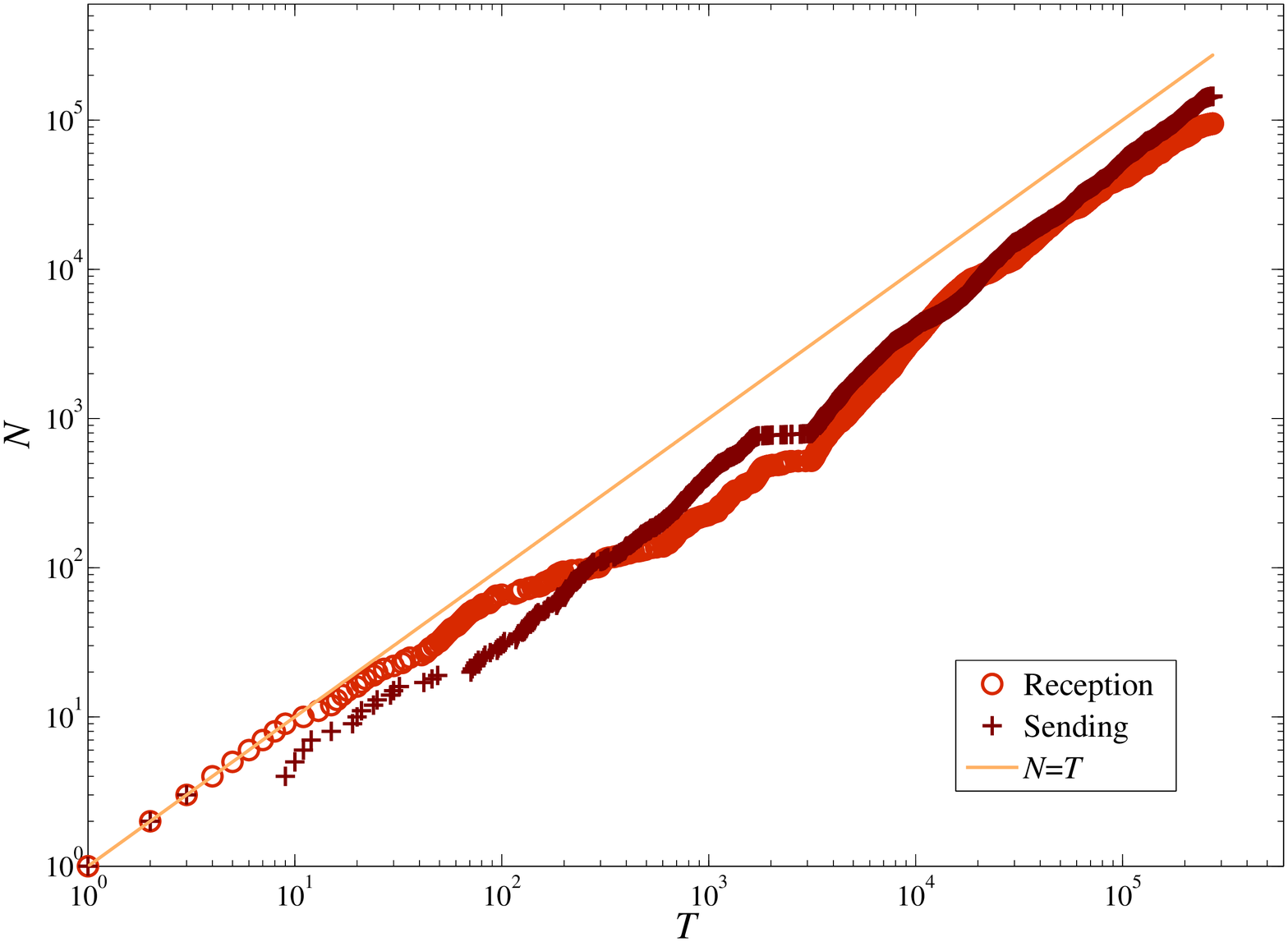}}
  \caption{Growth pattern of the number of different users $N$ with $T$
in \emph{Wealink}.}
\end{figure*}

\begin{figure*}
  \centerline{\includegraphics[height=2.5in]{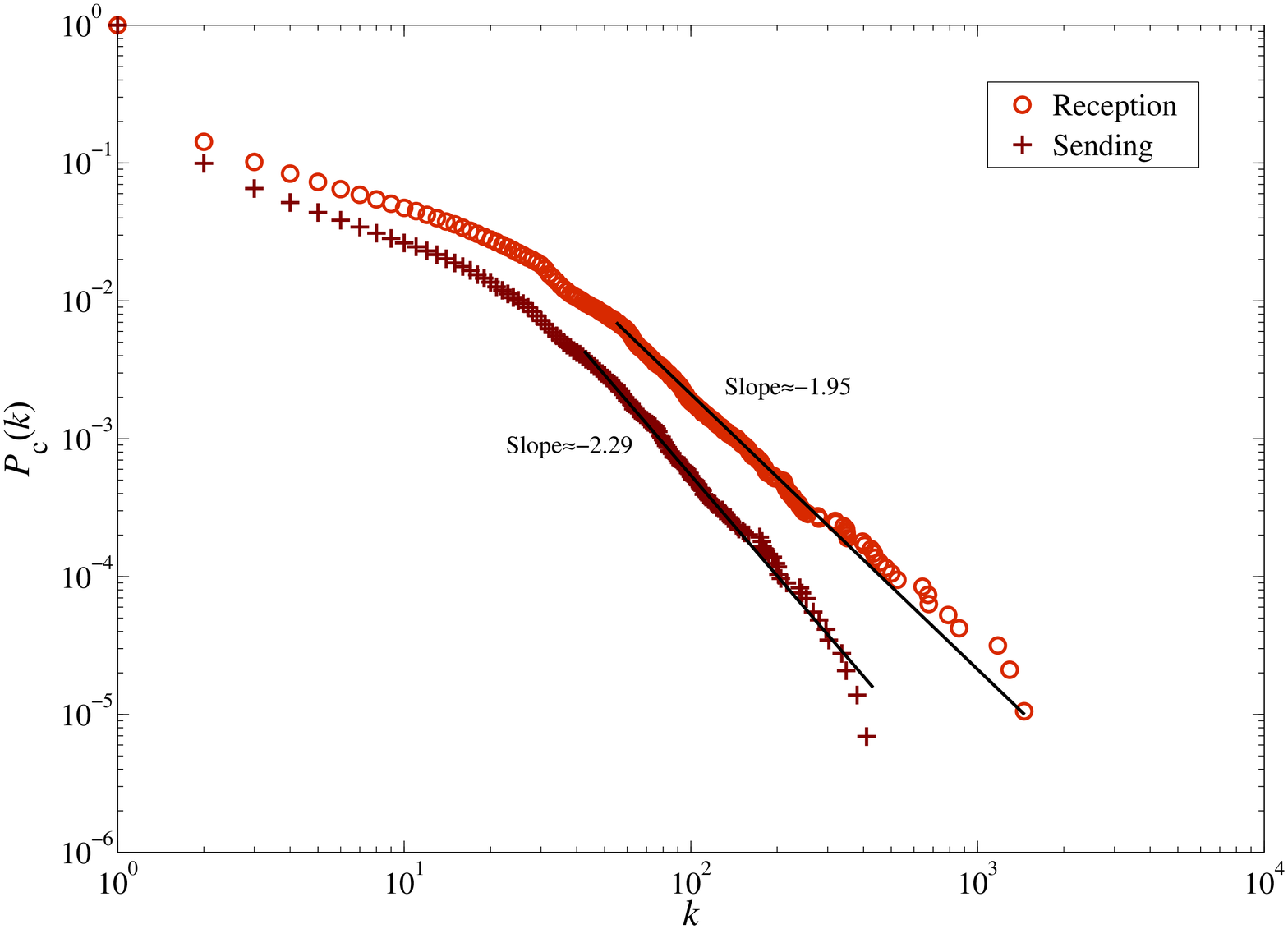}}
  \caption{The complementary cumulative distributions of the numbers
of sent and received invitations for users of \emph{Wealink}. Both
distributions have a power law tail with slope -2.29 for sent
invitations and -1.95 for received invitations.}
\end{figure*}

\section{Preferential linking}

The degree distribution of \emph{Wealink} shows power law features
[17]. This kind of distribution can be produced, as indicated by the
Barab\'{a}si-Albert (BA) model [26], through linear preferential
attachment, where new users tend to attach to already popular old
users. In \emph{Wealink}, as shown in Fig. 1, only when the sent
link invitations are accepted can the inviters and receivers become
friends and new edges appear in the social network. When new users
establish friend relationship with old users, or new edges are
established between old users, the old users with large degrees
could be preferentially selected.

To test the preference feature for different types of link
establishment, we separate the preferential linking into three
aspects: preferential acceptance, creation, and attachment.
Preferential acceptance implies that, the larger an old user's
degree is, the more likely he/she will accept link invitations from
other old users. Preferential creation implies that, the larger an
old user's degree is, the more likely his/her link invitations will
be accepted by the other old users. And preferential attachment
implies that new users tend to attach to already popular old users
with large degrees.

For instance, in Fig. 1, at time $T_6$, a new edge appears between
two old users $U_1$ and $U_2$. Old user $U_2$ who accepts a link
invitation can be chosen by preferential acceptance, and old user
$U_1$ who sends a link invitation can be chosen by preferential
creation. At time $T_8$, a new edge appears between old user $U_2$
and new user $U_5$, and old user $U_2$ can be chosen by preferential
attachment. Fig. 11 shows the relation between the degree $k$ of
users and the preference metric $\kappa$. We find that $\beta
\approx 1$ for preferential acceptance, creation, and attachment,
indicating linear preference.

\begin{figure*}
  \centerline{\includegraphics[height=2.5in]{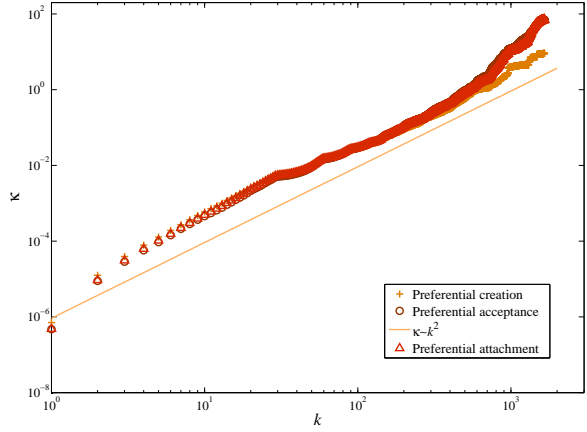}}
  \caption{Testing the preference feature for different types of link
establishment.}
\end{figure*}

The property of linear preference for the network can be generalized
to general OSNs. Mislove \emph{et al}. studied the evolution of
\emph{Flickr}; they defined preferential creation as a mechanism by
which users create new links in proportion to their outdegree, and
preferential reception as a mechanism where users receive new links
in proportion to their indegree [27]. They found that linear
preference holds for both cases, i.e. users tend to create and
receive links in proportion to their outdegree and indegree,
respectively. Leskovec \emph{et al}. studied the evolution of
\emph{Flickr}, \emph{del.icio.us}, \emph{Yahoo!Answers}, and
\emph{LinkedIn}, and examined whether the new users will
preferentially link to the old users with large degrees [24]. They
found that \emph{Flickr} and \emph{del.icio.us} show linear
preference, $\prod {(k)}  \sim k$, and \emph{Yahoo!Answers} shows
slightly sublinear preference, $\prod {(k)}  \sim {k^{0.9}}$.
\emph{LinkedIn} has a different pattern: for low degrees $k$, $\prod
{(k)}  \sim {k^{0.6}}$, and thus the preference is not obvious;
however, for large degrees, $\prod {(k)}  \sim {k^{1.2}}$,
indicating superlinear preference, i.e., the edges to higher degree
users are more sticky and high-degree users get super-preferential
treatment. Even though there are minor differences in the exponents
$\beta$ for different networks, we can say that $\beta \approx 1$,
implying that linear preference may be universal for OSNs.

According to this linear preference, we put forward a realistic
network model. Starting with a small network with $m_0$ nodes, at
every time step, there are two alternatives.

$A$. Growth and preferential attachment. With probability $p$, we
add a new node with $m_1$ ($<m_0$) edges that will be connected to
the nodes already present in the network based on the preferential
attachment rule of the BA model, i.e., the probability $\Pi $ that a
new node will be connected to old node $i$ with degree $k_i$ is
$\prod {({k_i})}  = {{{k_i}} \mathord{\left/
 {\vphantom {{{k_i}} {\sum\nolimits_j {{k_j}} }}} \right.
 \kern-\nulldelimiterspace} {\sum\nolimits_j {{k_j}} }}
$.

$B$. Preferential creation and acceptance. With probability $q=1-p$,
we add $m_2$ ($m_1+ m_2 \le m_0$) new edges connecting the old
nodes. The two endpoints of the edges are chosen according to linear
preference $\prod {({k_i})}  = {{{k_i}} \mathord{\left/
 {\vphantom {{{k_i}} {\sum\nolimits_j {{k_j}} }}} \right.
 \kern-\nulldelimiterspace} {\sum\nolimits_j {{k_j}} }}
$.

After $t$ time steps, the model leads to a network with average
number of nodes $\left\langle N \right\rangle  = m_0  + pt$. For
sparse real-world networks, $p>q$. When $p=1$, the model is reduced
to the traditional BA model. The model considers the introduction of
new nodes and new edges, which can be established either between new
nodes and old nodes or between old nodes. Most importantly, the
model integrates linear preference for acceptance, creation and
attachment found in the evolution process of real networks, and thus
captures realistic features of network growth.

The model has an analytic solution. Its stationary average degree
distribution for large $k$ is [28]
\begin{equation}
  P(k) \sim {k^{ - \frac{{3p{m_1} + 4q{m_2}}}{{p{m_1} + 2q{m_2}}}}},
  \label{4}
\end{equation}
showing a scale-free feature. According to Tab. 2, we obtain $p =
{p_{{\rm{Old - New}}}} + {p_{{\rm{New - Old}}}} =
{\rm{0}}{\rm{.7941}}$ and $q=0.1939$. The links created between two
new users are few and thus can be negligible. In addition,
$m_1=m_2=1$ for real growth of the network. Based on the parameters
and Eq. (4), we obtain $P(k)\sim k^ {-2.67}$. Fig. 12(a) shows the
numerical result which is obtained by averaging over 10 independent
realizations with $p ={\rm{0}}{\rm{.7941}}$ and $N$=223 482. Its
degree exponent 2.62 agrees well with the predicted value of 2.67.
Fig. 12(a) also presents the complementary cumulative degree
distribution of \emph{Wealink}, and we find that the predicted value
of the degree exponent 2.67 of the model achieves proper agreement
with the real value 2.91. The difference between real and
theoretical values may arise from the fact that $p$ and $q$ are
time-variant variables and not constants. Fig. 12(b) shows the
evolution of $p$ and $q$ and demonstrates the fact.

\begin{figure*}
  \centerline{\includegraphics[height=2.5in]{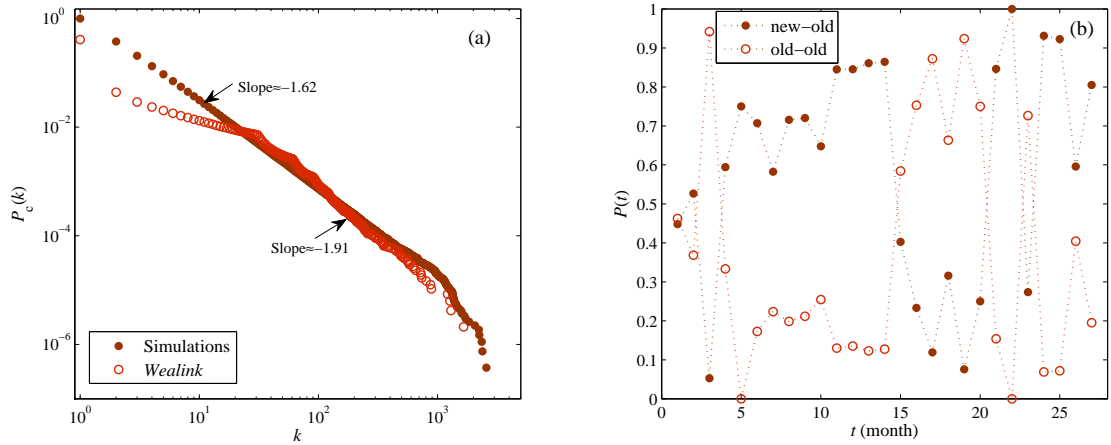}}
  \caption{(a) The complementary cumulative degree distributions of
\emph{Wealink} and the model. Both distributions have a power law
tail with slope -1.91 for \emph{Wealink} and -1.62 for the model.
(b) Evolution of the proportion of two kinds of edge. The Old-New
and New-Old types in Tab. 2 are integrated into the new-old type and
the old-old type still corresponds to the Old-Old type in Tab. 2.}
\end{figure*}

\section{Summary and discussion}
To conclude, we have unveiled the detailed growth of an OSN from a
microscopic perspective. Our study shows that the distribution of
intervals between sending and accepting link requests decays
approximatively exponentially, which is in obvious contrast to the
power law distribution of waiting time in emails, and there exists a
slightly negative correlation between reciprocation time and degrees
of inviters/accepters. The distributions of intervals of user
behaviors, such as sending or accepting link requests, follow a
power law with a universal exponent, indicating the bursty nature of
the user act. We finally study the preference phenomena of the OSN
and find that for preferential selection linear preference holds for
preferential sending and reception, and for preferential linking
linear preference also holds for preferential acceptance, creation
and attachment. We propose a network model which captures real
features of network growth and can reproduce the degree distribution
of the OSN.

It is noteworthy that, although there is a close relation between
the microscopic growth of networks and global network structure or
structural metric evolution, it is still quite hard to bridge the
gap between macro and micro perspectives of OSNs. For instance,
preferential linking may possibly supply some information on the
degree distribution of networks; however, it may not tell us much
about the other properties of networks, such as clustering or
community structure. Thus to gain an in-depth comprehension of OSNs,
other microscopic behaviors of users, such as homophily, need to be
studied in detail; a complementary research framework integrating
macro and micro perspectives will also be indispensable.


\begin{thebibliography}{10}

\bibitem{1.}
T. O'Reilly, What is Web 2.0,
http://oreilly.com/web2/archive/what-is-web-20.html.

\bibitem{2.}
Y.Y. Ahn, S. Han, H. Kwak, S. Moon, H. Jeong, Analysis of
topological characteristics of huge online social networking
services, in: Proceedings of the 16th international conference on
World Wide Web, ACM Press, New York, 2007, pp. 835-844.

\bibitem{3.}
M. Torkjazi, R. Rejaie, W. Willinger, Hot today, gone tomorrow: on
the migration of MySpace users, in: Proceedings of the 2nd ACM
workshop on Online social networks, ACM Press, New York, 2009, pp.
43-48.

\bibitem{4.}
S.A. Golder, D. Wilkinson, B.A. Huberman, arXiv:cs/0611137.

\bibitem{5.}
K. Lewis, J. Kaufman, M. Gonzalez, A. Wimmer, N. Christakis, Social
Networks 30 (2008) 330-342.

\bibitem{6.}
B. Viswanath, A. Mislove, M. Cha, K.P. Gummadi, On the evolution of
user interaction in Facebook, in: Proceedings of the 2nd ACM
workshop on Online social networks, ACM Press, New York, 2009, pp.
37-42.

\bibitem{7.}
J.P. Onnela, F. Reed-Tsochas, Proc. Natl. Acad. Sci. USA 107 (2010)
18375-18380.

\bibitem{8.}
A. Mislove, M. Marcon, K.P. Gummadi, P. Druschel, B. Bhattacharjee,
Measurement and analysis of online social networks, in: Proceedings
of the 7th ACM SIGCOMM conference on Internet measurement, ACM
Press, New York, 2007, pp. 29-42.

\bibitem{9.}
B. Wellman, Science 293 (2001) 2031-2034.

\bibitem{10.}
C. Licoppe, Z. Smoreda, Social Networks 27 (2005) 317-335.

\bibitem{11.}
W.S. Bainbridge, Science 317 (2007) 472-476.

\bibitem{12.}
N. Shadbolt, T. Berners-Lee, Sci. Am. 299(4) (2008) 76-81.

\bibitem{13.}
L. L\"{u}, T. Zhou, Physica A 390 (2011) 1150-1170.

\bibitem{14.}
R. Dunbar, Behavioral and Brain Sciences 16 (1993) 681-735.

\bibitem{15.}
P. Holme, C.R. Edling, F. Liljeros, Social Networks 26 (2004)
155-174.

\bibitem{16.}
M.E.J. Newman, Phys. Rev. E 67 (2003) 026126.

\bibitem{17.}
H. Hu, X. Wang, Phys. Lett. A 373 (2009) 1105-1110.

\bibitem{18.}
H.-B. Hu, X.-F. Wang, EPL 86 (2009) 18003.

\bibitem{19.}
H. Chun, H. Kwak, Y.H. Eom, Y.Y. Ahn, S. Moon, H. Jeong, Comparison
of online social relations in terms of volume vs. interaction: A
case study of Cyworld, in: Proceedings of the 8th ACM SIGCOMM
conference on Internet measurement, ACM Press, New York, 2008, pp.
57-70.

\bibitem{20.}
M. Szell, S. Thurner, Social Networks 32 (2010) 313-329.

\bibitem{21.}
A.-L. Barab\'{a}si, Nature 435 (2005) 207-211.

\bibitem{22.}
A. V\'{a}zquez, J.G. Oliveira, Z. Dezs\"{o}, K.-I. Goh, I. Kondor,
A.-L. Barab\'{a}si, Phys. Rev. E 73 (2006) 036127.

\bibitem{23.}
H. Jeong, Z. N\'{e}da, A.-L. Barab\'{a}si, Europhys. Lett. 61 (2003)
567-572.

\bibitem{24.}
J. Leskovec, L. Backstrom, R. Kumar, A. Tomkins, Microscopic
evolution of social networks, in: Proceeding of the 14th ACM SIGKDD
international conference on Knowledge discovery and data mining, ACM
Press, New York, 2008, pp. 462-470.

\bibitem{25.}
H.A. Simon, Biometrika 42 (1955) 425-440.

\bibitem{26.}
A.-L. Barab\'{a}si, R. Albert, Science 286 (1999) 509-512.

\bibitem{27.}
A. Mislove, H.S. Koppula, K.P. Gummadi, P. Druschel, B.
Bhattacharjee, Growth of the Flickr social network, in: Proceedings
of the first workshop on Online social networks, ACM Press, New
York, 2008, pp. 25-30.

\bibitem{28.}
J.-L. Guo, An evolution model for forum networks, preprint.

\end{thebibliography}
\end{document}